\documentclass[twocolumn,superscriptaddress,showpacs,preprintnumbers,amsmath,amssymb]{revtex4}
\usepackage{graphicx}
\usepackage{dcolumn}
\usepackage{bm}
\begin{document}
\title{Geometric quantum gates  robust against stochastic control errors}
\author{Shi-Liang Zhu}
\affiliation{Institute for Scientific Interchange Foundation,
Viale Settimio Severo 65, I-10133 Torino, Italy}
\affiliation{School of Physics and Telecommunication Engineering,
South China Normal University, Guangzhou, China}
\author{Paolo Zanardi}
\affiliation{Institute for Scientific Interchange Foundation,
Viale Settimio Severo 65, I-10133 Torino, Italy}
\begin{abstract}

We analyze a scheme for quantum computation where quantum gates
can be continuously changed from  standard dynamic gates to purely
geometric ones. These gates  are enacted by controlling a set of
parameters that  are subject to unwanted stochastic fluctuations.
This kind of noise results in a departure from the ideal case that
can be quantified by a  gate fidelity.  We find that the maximum
of this fidelity corresponds to quantum gates with  a vanishing
dynamical phase.

\end{abstract}

\maketitle
\newpage

An essential prerequisite for quantum computation is the ability
of maintaining quantum coherence and quantum entanglement in a
information-processing  system \cite{qip}. Unfortunately, since
both these properties are very fragile against control errors as
well as against unwanted couplings with environment, this goal
is  extremely hard to achieve. To this end several strategies have
been developed, most notably: quantum error correction\cite{Shor},
decoherence-free subspace\cite{Duan97}, and bang-bang
techniques\cite{Viola}.

Quantum computation implemented by geometric
phases\cite{Berry,Aharonov,Zhu_PRL2000} is believed to be another
approach which can be used to overcome certain kinds of
errors \cite{Zanardi,Jones,Duan,Wang,Zhu_PRL2002,Zhu_PRA2003,Leibfried,Solinas}.
It has been shown a universal set of quantum gates can be realized
by geometric phases. However, the statement that quantum gates
achieved by this way may have built-in fault-tolerant features
(due to the fact that geometric phases depend only on some global
geometric properties) has still the status of a conjecture and it has been the subject
of some debate in the literature.
Indeed this alleged resilience against errors of geometrical gates  has been
doubted by some numerical calculations with certain decohering
mechanisms\cite{Nazir,Blais}. On the other hand, analytical
results show  that adiabatic Berry's phase itself may be robust
against dephasing\cite{Carollo} and stochastic fluctuations of control parameters
\cite{Chiara}. These latter  provide a sort of  indirect evidence of
 the robustness of adiabatic geometric quantum computation;
 a more direct evidence would be given by a  comparison  between the fidelity
of geometric gates and standard dynamic gates in presence of an
error source. So far, this kind of convincing evidence, clearly
showing that geometric quantum computation is robust against some
realistic noise sources, is still missing.

In this paper we shall consider a parametric family of quantum
gates  subject to stochastic fluctuations of the control
parameters.  The departure from the ideal (i.e., no-fluctuation)
case can be quantified by gate fidelity. We compute  such a
fidelity  in a model where quantum gates can continuously change
from standard dynamic gates to purely geometric ones. We find that
{\em the maximum of fidelity corresponds to those cases in which
the dynamical phase accumulated  over the gate operation is zero}.
This  provides a clear evidence of the robustness of nonadiabatic
geometric computation in this specific scheme. The predictions
presented here may be experimentally tested in some quantum
computer prototypes. Moreover, this robustness of nonadiabatic
geometric computation is relevant to  experiments on quantum
information processing; indeed it is generally believed that the
control parameter nonuniformity is one of the most dangerous
sources of errors for qubits in solid-state systems
\cite{Makhlin}, NMR \cite{Vandersypen} and trapped ions
\cite{Garcia-Ripoll}, etc..

{\sl Quantum computation via a pair of orthogonal cyclic states}--
For universal quantum computation, it is sufficient to enact two
of noncommuting single-qubit gates and one nontrivial two-qubit
gate. Before to study the fidelity of a quantum gate subject to
noise, we recall a scheme to implement a universal set of quantum
gates by using a pair of orthogonal cyclic
states\cite{Zhu_PRL2002,Zhu_PRA2003}. A single qubit gate given by

$$U^{(1)}=\left (
\begin{array}{ll}
e^{i\gamma}\cos^2\frac{\chi}{2}+e^{-i\gamma}\sin^2\frac{\chi}{2}
& i\sin\chi \sin\gamma  \\
i \sin\chi \sin\gamma &
e^{i\gamma}\sin^2\frac{\chi}{2}+e^{-i\gamma}\cos^2\frac{\chi}{2}
\end{array}
\right )
$$
can be obtained  when a pair of cyclic states $|\pm\rangle$ can be
found for  a unitary operation $U^{(1)}$, i.e.,
$U^{(1)}|\pm\rangle=e^{\pm i \gamma} |\pm\rangle.$
 Here $\chi$ is related to the initial cyclic states,
and $\gamma$ is a phase accumulated in the gate evolution. Usually
the total phase $\gamma$ consists of both geometric ($\gamma_g$)
and dynamic components ($\gamma_d$), and $U$ is specified as a
geometric gate if $\gamma$ is a pure geometric phase. Moreover, a
conditional two-qubit gate can also be implemented if there exist, conditional
to the state of control qubit,
two different pairs of cyclic states of the  target qubit.
In terms of the computational basis
$\{ |00\rangle, |01\rangle, |10\rangle, |11\rangle \}$, where the
first (second) bit represent the state of  the control (target)
qubit. The unitary operator describing  the conditional two-qubit
gate is given by
$U^{(2)}= \rm{diag}(U_{(\gamma^0,\chi^0)},U_{(\gamma^1,\chi^1)})$, under
the condition that the control qubit is off resonance in the
manipulation of the target qubit. Here $\gamma^\delta$
($\chi^\delta$) represents the total phase (the cyclic initial
state) of the target qubit when the control qubit is in the state
$\delta (=0,1)$. This scheme can be implemented in several
realistic physical systems\cite{Zhu_PRL2002,Zhu_PRA2003}.

{\sl Control parameter fluctuation}--
 A simple approach for implementing  $U^{(1)}$ and $U^{(2)}$ is to use an effective rotating
magnetic field to manipulate the state of qubits. In this case,
the Hamiltonian for single qubit reads
\begin{equation}
\label{rotated-field} H=(\omega_0\sigma_x \cos\omega t+\omega_0
\sigma_y \sin\omega t+\omega_1 \sigma_z)/2,
\end{equation}
where $\omega_i=-g\mu B_i/\hbar$ $(i=0,1)$ with $g$ $(\mu)$ being
the gyromagnetic ratio (Bohr magneton), and $B_i$ $(i=1,2)$ acts
as an external controllable parameter, and its magnitude can be
experimentally changed. We also use this Hamiltonian to manipulate
the target qubit in the implementation of two qubit gate when the
control qubit is off resonance. The Hamiltonian of the two-qubit
system is given by the Hamiltonian (\ref{rotated-field}) plus the
coupling Hamiltonian acting on the two qubits. In the non-ideal case
the control fields contain randomly fluctuating components, here we assume
that $\omega_i$ is flatly distributed in the interval
$[(1-\delta_i)\omega_i,(1+\delta_i)\omega_i]$ with $\delta_{i}$ a
constant, and then we numerically calculate the average fidelity of
quantum gates $U^{(1)}$ and $U^{(2)}$. But we will assume that
$\omega$ is not affected by random fluctuations; this seems  reasonable since
frequency may be well controlled in a realistic experiment.

The average fidelity of a
quantum gate we study is defined by
\begin{equation}
\label{Fidelity} \overline{F}=
\lim_{N\to\infty}\frac{1}{N}\sum\limits_{j=1}^{N} \overline{F_j}=
\lim_{N\to\infty}\frac{1}{N}\sum\limits_{j=1}^{N}\overline{\left|\langle\psi^{in}_{j}
|\hat{U}^{\dagger}_{id} |\psi^{out}_{j}\rangle \right|^2},
\end{equation}
where $|\psi^{in}_{j}\rangle=[\cos\frac{\theta_{j}}{2}
e^{-i\varphi_j/2},\sin\frac{\theta_{j}}{2} e^{i\varphi_j/2}]^T$ or
$|\psi^{in}_{j}\rangle=[-\sin\frac{\theta_{j}}{2}
e^{-i\varphi_j/2},\cos\frac{\theta_{j}}{2} e^{i\varphi_j/2}]^T$
($T$ denotes the matrix  transposition) is an input state.
$\theta_j\in [0,\pi]$ and $\varphi_j \in [0,2\pi]$ are randomly
chosen in our numerical calculation. $U_{id}$ is the ideal quantum
gate without any control parameter fluctuation  and
$|\psi^{out}_{j}\rangle$ is the output state after a noisy
gate operation  when the input state is $|\psi^{in}_{j}\rangle$.
In the numerical calculation, we randomly choose one input state
$|\psi^{in}_j\rangle$, and then calculate the average fidelity of
$\overline{F_j},$ up to  satisfactory convergence, for $M$
configurations of  fluctuations of magnetic fields. After that, we
randomly choose the next input state and repeat the above
calculation until deriving the fidelity of this specific input
state with satisfactory convergence. We repeat $N$ times to
calculate the average fidelity by randomly choosing
$|\psi_j^{in}\rangle$. In our numerical calculations below, we get
small statistical errors when $M$ and $N$ are about several
hundreds to one thousand.

{\sl Fidelity of single-qubit gates}-- The Schr\"{o}dinger
equation with Hamiltonian (\ref{rotated-field}) can be
analytically solved, and single qubit gates $U^{(1)}$ can be
achieved, where $\chi=\arctan[\omega_0/(\omega_1-\omega)]$ is the
angle between the initial state and the symmetric axis of the
rotating field. The corresponding phases for one cycle are given
by
$\gamma_d=-\pi[\omega_0^2+\omega_1(\omega_1-\omega)]/\omega\Omega,$
$\gamma_g = -\pi[1-(\omega_1-\omega)/\Omega]$, and $\gamma =
-\pi(1+\Omega/\omega)$ with
$\Omega=\sqrt{\omega_0^2+(\omega_1-\omega)^2}$. We can choose any
two processes with different values
$\{\omega^{j},\omega_0^{j},\omega_1^{j} \}$ $(j=1,2)$ satisfying
the constrain $\sin\gamma_1 \sin\gamma_2
\sin(\chi_2-\chi_1)\not=0$ to enact  two noncommuting single-qubit
gates\cite{Zhu_PRA2003}.

What is  remarkable here is that quantum gate $U^{(1)}$ implemented in this
way can be varied continuously from a standard dynamic gate to a pure
geometric gate, by changing the external parameters
$\{\omega,\omega_0,\omega_1 \}$. Hence, this scheme looks like an ideal
model to compare the difference between geometric quantum gates
and standard dynamic gates. It is straightforward to verify from
the expression of $\gamma_d$ that the dynamic phase is zero under
the condition $\omega=(\omega_0^2+\omega_1^2)/\omega_1.$ Thus we
can obtain  purely  geometric gates by choosing these specific
parameters, such that  $\gamma_d=0$ in the whole process.
It has been shown that $\chi$ in $U^{(1)}$ can be controlled
independently by the symmetric axis of the rotating field, and a
specific quantum gate can be realized  by a fixed phase $\gamma$.
For example, the Hadamard gate is obtained  by
$\gamma=(n_1+1/2)\pi$ and $\chi=(n_2+1/4)\pi$ (with $n_{1,2}$
integers). Therefore, we assume the total phase is fixed, for
concreteness, we choose $\gamma=-\beta\pi$ with $\beta$ a
constant. It is straightforward to check that $\omega=(\omega_1
\pm \sqrt{\omega_1^2-\eta(\omega_0^2+\omega_1^2)})/\eta$ with
$\eta=2\beta-\beta^2$ guarantee $\gamma=-\beta \pi$.
The obvious requirement of $\omega$ reality  implies $\eta\omega_0^2 \leq
(1-\eta)\omega_1^2$. On the other hand, the dynamic phase is zero
under the condition
\begin{equation}
\label{G_phase1} \omega_1=\omega_0 \sqrt{\eta/(1-\eta)}.
\end{equation}

\begin{figure}[htbp]
\centering \vspace{-3cm}
\includegraphics[height=10cm,width=8cm]{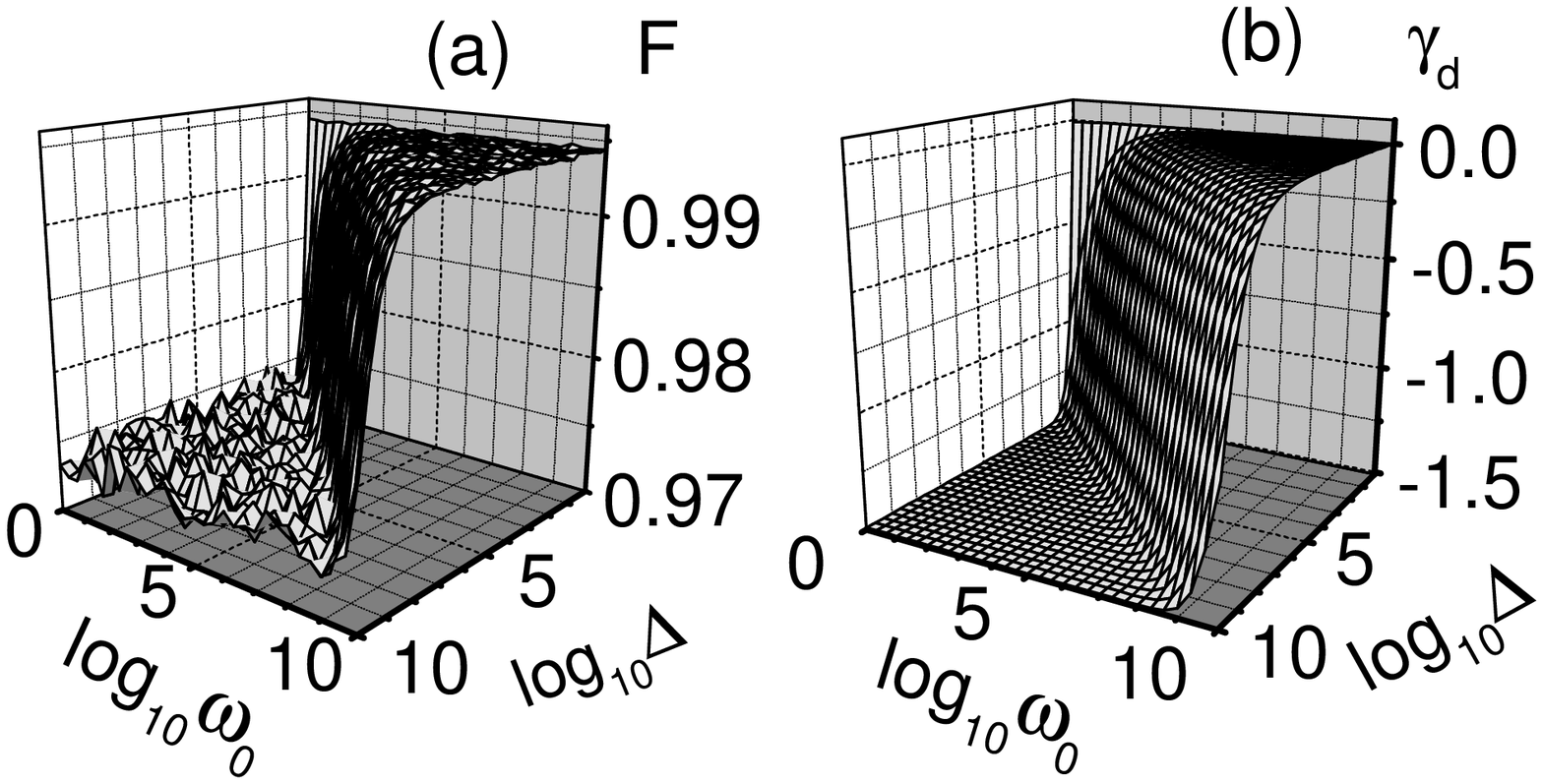}
\vspace{-3.2cm} \caption{The fidelity and phase in single qubit
gates. (a) Fidelity, (b) dynamic phase.} \label{Fig1}
\end{figure}

\begin{figure}[htbp]
\centering
\includegraphics[height=4cm,width=8cm]{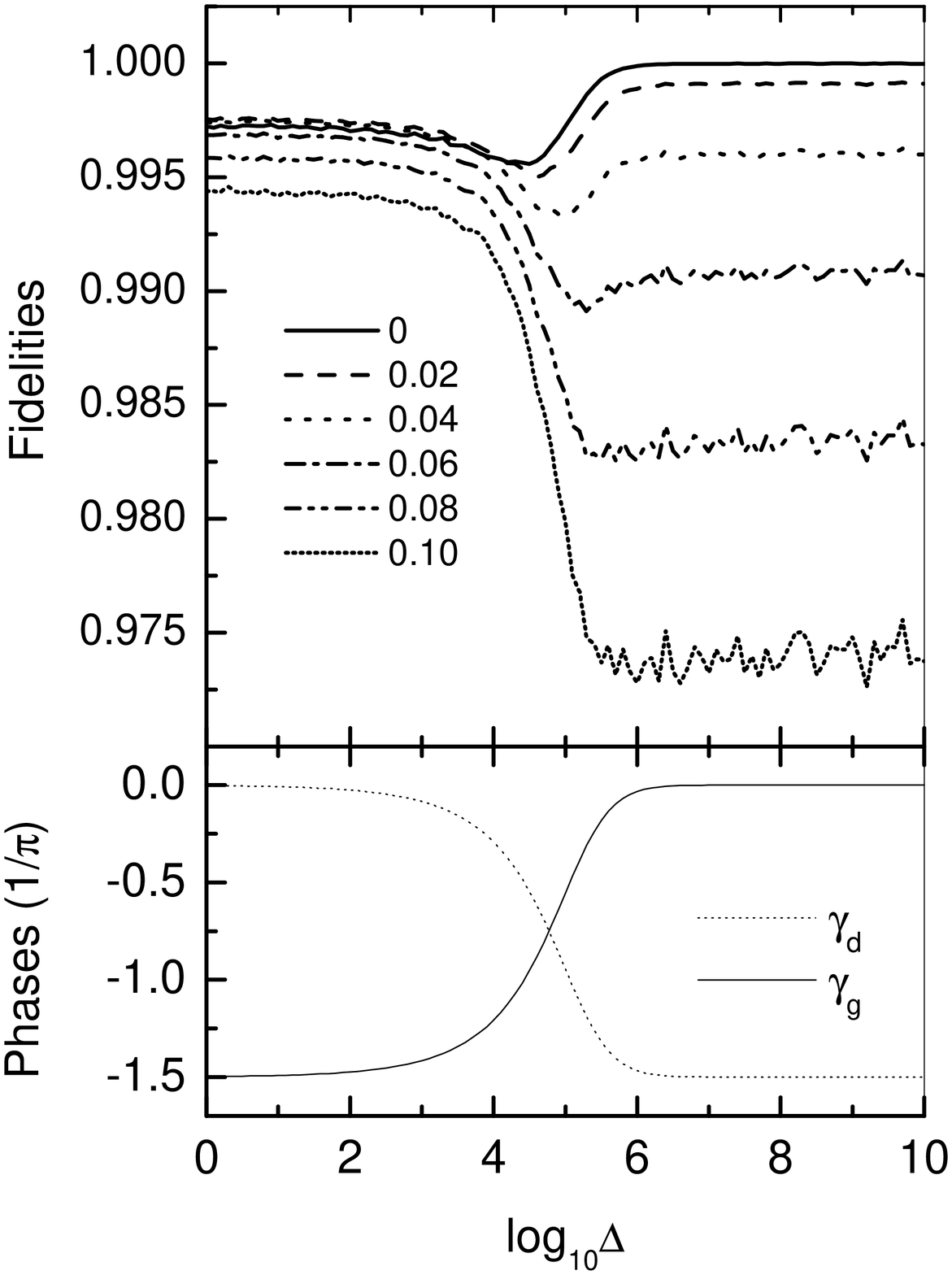}
\caption{The fidelities and phases in single qubit gate for
$\delta_0=0.1$. The values of $\delta_1$ are indicated. (a)
Fidelities, (b) dynamic and geometric phases.} \label{Fig2}
\end{figure}

The fidelity $F(\omega_0,\Delta)$ and dynamic phase of this gate
for typical parameters are shown in Fig.1. Here $\Delta$ is
defined by $\omega_1=\sqrt{\eta/(1-\eta)}\omega_0+\Delta$.
$\beta=3/2$ is chosen just as an  example;  we have checked that
the main features described here are independent on this
parameter. We plot $F(\omega_0,\Delta)$ instead of
$F(\omega_0,\omega_1)$ to guarantee that $\omega$
$(=2(2\omega_1-\sqrt{\omega_1^2-3\omega_0^2})/3)$ is a real number
in the whole region. We can see several remarkable features from
Fig.\ref{Fig1}, where $\delta_0=\delta_1=0.1$: (i) the maximum of
fidelity is along the line described by $\Delta=0$, where the
dynamic phase is zero; (ii) Fidelity is monotonically  decreasing
by  increasing $\Delta$. The change of fidelity is slow at small
$\Delta$, but very sharp near $\Delta \sim \omega_0$, and it is
also slow for large values  of $\Delta$. Remarkably, the changes
of the absolute value of dynamic phases with $\Delta$ are just as
the same as the changes of the average fidelity of gates.
Therefore, it clearly shows the close relation between the
fidelity of quantum gates and the component of the dynamic
(geometric) phase. This feature will  be  demonstrated also in the
two-qubit case addressed later. We also checked  the fidelity for
different $\delta_0$ and $\delta_1$. The fidelity and quantum
phases as a function of $\Delta$ for $\omega_0=10^5$ are shown in
Fig.\ref{Fig2}. We observe that the two main features discussed
above appear, for the case $\delta_0=0.1,$ when $\delta_1$ is
greater than $0.04.$
 However, when  $\delta_1$ is less than $0.04$, it is worth pointing out that
the fidelity in the  points with  $\Delta=0$ is a local maximum,
since there is a dip clearly shown nearby $\Delta \sim \omega_0;$
the largest fidelity appears when dynamic phase is dominant. We
also numerically computed  the fidelity for fixed $\delta_1$ but
varied $\delta_0$ (not shown), the main features for different
$\delta_0$ are totally  similar to  those in Fig.\ref{Fig1}. The
fact that the fidelity is larger when $\Delta$ ($\omega_1$) is
large but with small fluctuation of $\omega_1$ can be
qualitatively explained from the dynamic phase $\gamma_d.$
The deviation of $\gamma_d$ from the noiseless case is
dominated by the fluctuations of $\omega_1$ when $\Delta$ is much
larger than $\omega_0.$ Therefore the infidelity should be small if the
fluctuations of $\omega_1$ are very small.

{\sl Fidelity of two-qubit gates}-- We now numerically compute the
fidelity of a two-qubit gate. We assume that the qubit-qubit
interaction is given by $ H_I=J\sigma_z^{(1)}\sigma_z^{(2)}/2$.
This kind of coupling between qubits can be naturally realized,
e.g., in quantum computer models with NMR and superconducting
charge qubits coupled through capacitors. When the target qubit is
manipulated by a rotating field described by
Eq.(\ref{rotated-field}) and control qubit is off resonance, it is
shown that a two-qubit gate $U^{(2)}$ with
 $\chi^{\delta}=\arctan
[\omega_0/(\omega_1^\delta-\omega)]$ and $\gamma^\delta   =
-\pi(1+\Omega^\delta/\omega)$ can be implemented. Here
$\omega_1^\delta=\omega_1+(2\delta-1)J$ and
$\Omega^\delta=\sqrt{\omega_0^2+(\omega_1^\delta-\omega)^2}$,
$\omega$, $\omega_0$, and $\omega_1$ are parameters for the target
qubit.
 The corresponding phases for one cycle are given by $\gamma_d^\delta =
-\pi[\omega_0^2+\omega_1^\delta
(\omega_1^\delta-\omega)]/\omega\Omega^\delta,$ and
$\gamma^\delta_g =
-\pi[1-(\omega_1^\delta-\omega)/\Omega^\delta]$. Besides, it is
easy to check from $\gamma_d^\delta =0$ that the geometric
two-qubit gates are realized whenever $\omega = 2\omega_1,$ and $
\omega_1^2 = \omega_0^2+J^2$\cite{Zhu_PRA2003}.

\begin{figure}[htbp]
\centering \vspace{-3cm}
\includegraphics[height=10cm,width=8cm]{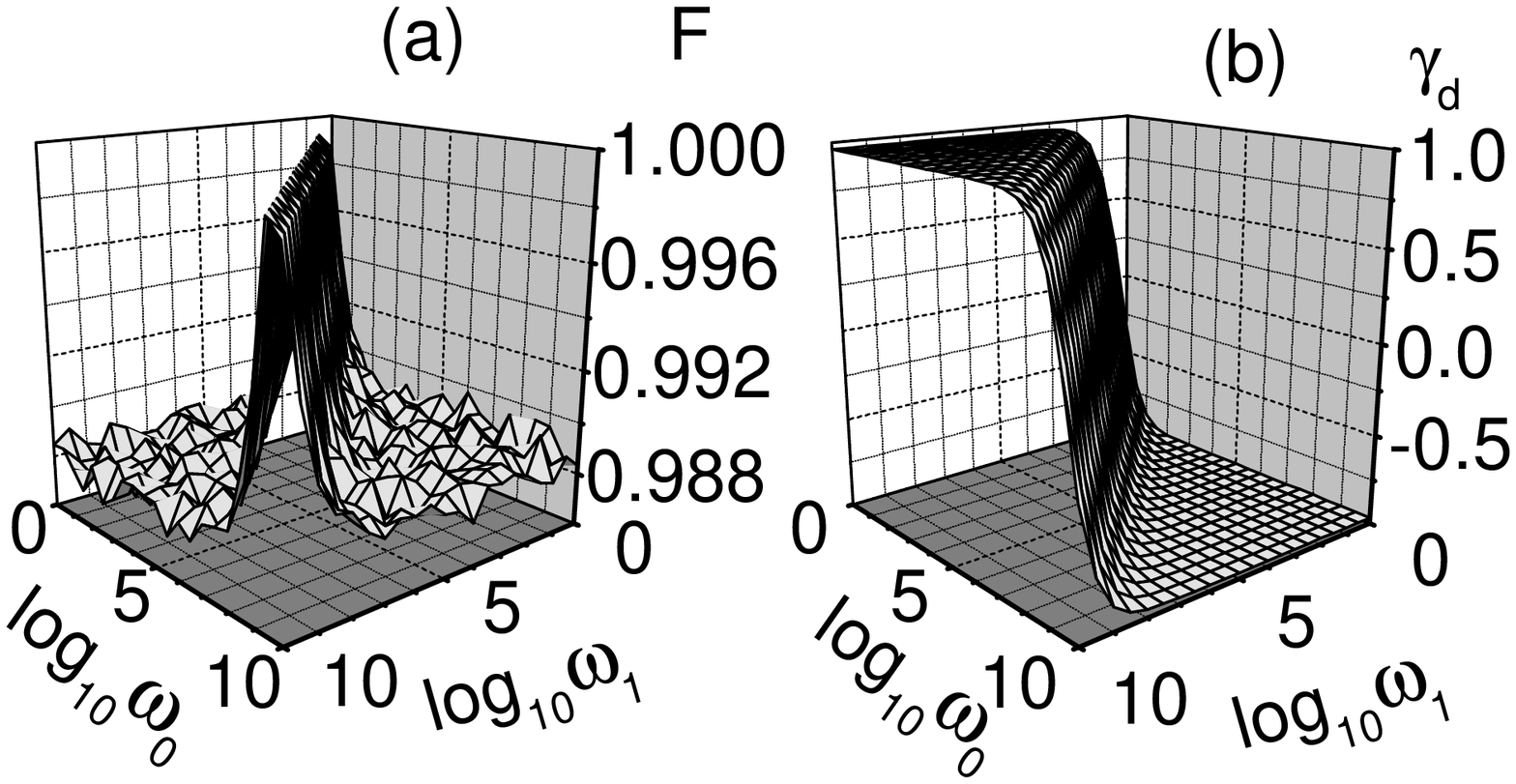}
\vspace{-3.2cm} \caption{The fidelity and phase in two-qubit gates
for $\delta_0=\delta_1=0.1$, $\delta=0$ and $\alpha=\sqrt{3}$. (a)
Fidelity, (b) dynamic phase.} \label{Fig3}
\end{figure}

There is a lot of freedom in  choosing parameters to implement a
geometric quantum gate in the present scheme. One possible choice is
given by  $ \omega =\omega_1+\sqrt{1+\alpha^2}\omega_0$ and $
J=\alpha\omega_0$, thus the speed of the purely geometric gate is of the same
order of that of the  dynamic gate. To see the relation between the
fidelity and the quantum phase, we plot the fidelity and dynamic phase
of gate $U^{(2)}$ just when $\delta=0$ as a function of the
external parameters $\omega_0$ and $\omega_1$ in Fig.3. In this
case, the dynamic phases change sharply nearby the line described
by $\log_{10}\omega_1=\log_{10}(\sqrt{1+\alpha^2}\omega_0)$, where
dynamic phases are zero, since it is straightforward to find that
under the condition
\begin{equation}
\label{Zero_dynamic} \omega_1=\sqrt{1+\alpha^2}\omega_0,
\end{equation}
$\gamma_d^\delta$ is zero. We can see from Fig.\ref{Fig3} that the
maximum of fidelity is along the line where the dynamic phase is
zero. Moreover, compared with the single-qubit case shown in
Fig.\ref{Fig1}, the fidelity of the gate shown here decreases
quickly, since the dynamic phase changes sharply when the
parameters do not satisfy  Eq.(\ref{Zero_dynamic}). Therefore,
it is clearly shown the close relation between the change of
fidelity and the change of dynamic component of the phase.

\begin{figure}[htbp]
\centering
\includegraphics[height=5cm,width=8cm]{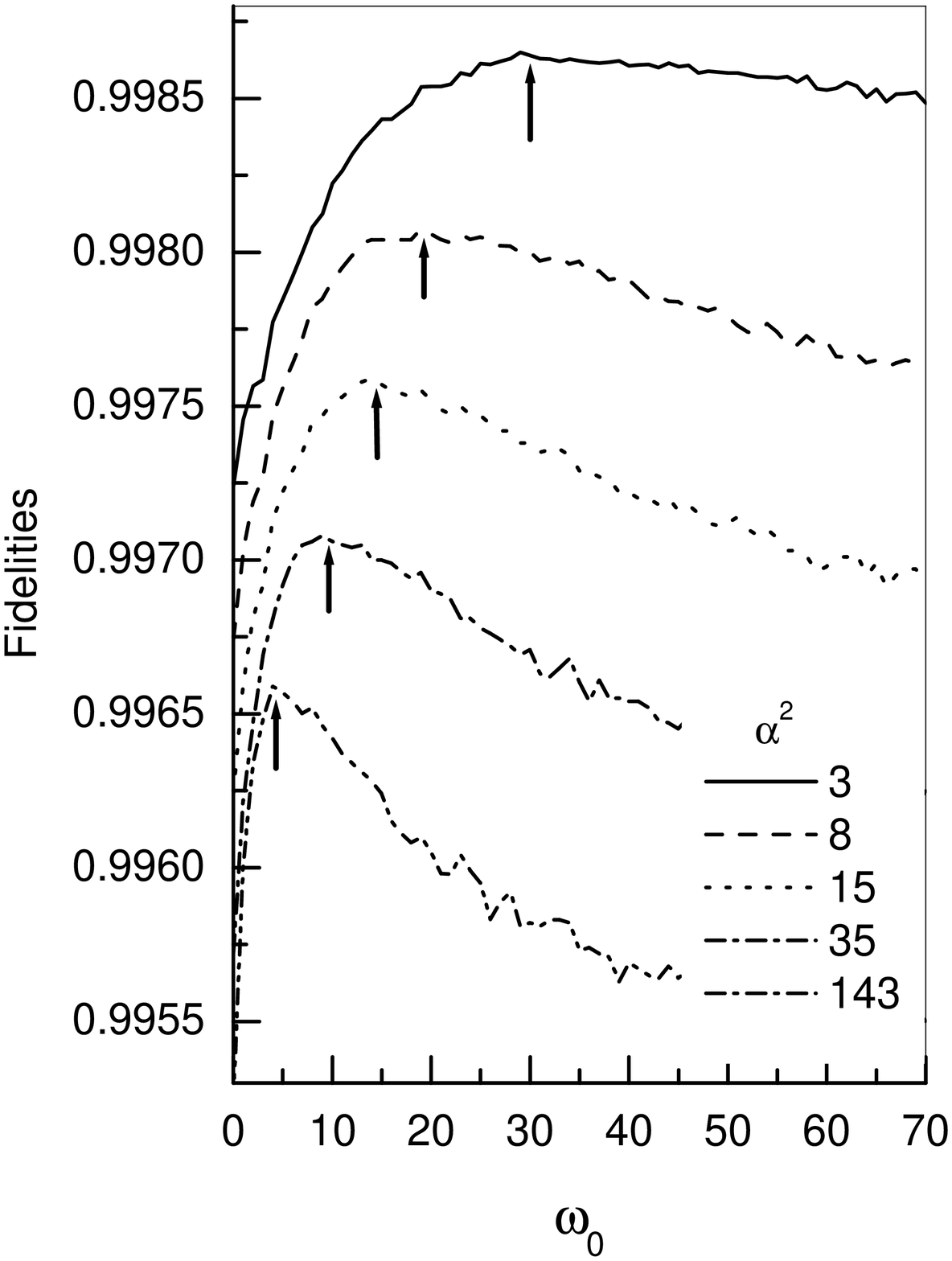}
\caption{The fidelities of two-qubit gates. The curves for
different $\alpha$ are vertically shifted for clarity. Points with
zero dynamic phases shown in Eq.(\ref{Zero_dynamic}) are denoted
by arrows.} \label{Fig4}
\end{figure}

To  show  in a more clear fashion that the maximum of fidelity is
along the line described by Eq.(\ref{Zero_dynamic}) and that this
feature is independent on the  parameter $\alpha$, we numerically
computed  the average fidelity of $U^{(2)}$ when the state of the
control qubit is unfixed. The fidelity as a function of $\omega_0$
for $\omega_1=60$, $\delta_0=\delta_1=0.05$ and
$\alpha=\sqrt{3},\sqrt{8},\sqrt{15},\sqrt{35},\sqrt{143}$ are
plotted in Fig.\ref{Fig4}. It is easy to derive from
Eq.(\ref{Zero_dynamic}) that the dynamic phase is zero at
$\omega_0=30,20,15,10,5$, and these points are denoted by arrows
in Fig.\ref{Fig4}. We observe that the maxima of fidelity are
indeed at the points described by Eq.(\ref{Zero_dynamic}), where
the dynamic phases are zero; this  property is independent on
$\alpha$.

{\sl Comparison with previous results}-- We would like to compare
the results here with the previous results in literature.
It has been  shown that the effects of fluctuations of the control
parameters\cite{Blais} or decoherence described by Lindblad
form\cite{Nazir} on nonadiabatic gates are more severe than for
the standard dynamic gates. We note that in
Refs.\cite{Blais,Nazir} the geometric gate is implemented by three
rotations, but only one operation is used to realize a standard
dynamic gate. Thus  a direct comparison is somewhat not appropriate,
 and one can not rule out other possibilities. In the present model,
however, the operations for both dynamic gate and geometric gate
are totally the same, except that the controllable parameters vary
continuously. In this sense  this model looks definitely more suitable
to assess  the difference between geometric gates and dynamic
gates as far as noise resilience is concerned.

Before concluding, we would like to point out that the scheme
studied here is based on nonadiabatic operations. Note that
$\omega$ is of the same order of the magnitude as $\omega_0$ or
$\omega_1$ in both one and two-qubit gate operations. This implies
that the speed of geometric quantum gate  here investigated  is
comparable with that of the dynamic quantum gate. In contrast, the
speed of quantum gate based on adiabatic Berry's phase is much
lower than that of gate using dynamic phase, since the adiabatic
condition requires that both $\omega_0$ and $\omega_1$ should be
much larger than $\omega$. Thus the speed constraint required by
adiabatic geometric gates is removed in the present scheme, and
the results discussed above  show the robustness of this
nonadiabatic geometric gates against noise in the control
parameters. This considerations suggest that this approach to
quantum computation should be significant in quantum information
processing.

{\sl Conclusions}--
In this paper  we have numerically computed  the average fidelity
of a family of quantum gates, which vary continuously from the standard
dynamic gates to pure geometric gates with the change of
experimentally controllable parameters. We found that the maximum
of gate fidelity is achieved  along the line where the dynamical
phase is zero. We believe that the results presented in this paper
provide a first convincing evidence of the robustness of geometric
computation. Our predictions can be, in principle, experimentally
tested   in already existing quantum computer prototypes, such as
NMR, trapped ions and superconducting qubits, etc..

This work was supported by the European Union project TOPQIP
(contract IST-2001-39215). S.L.Z. was supported by the NSFC under
Grant No.10204008 and the NSF of Guangdong under Grant No.021088.

\end{document}